\documentstyle[12pt]{article}
 
\title{PARTON DISTRIBUTION WITHIN VIRTUAL PHOTON
AND DIFFRACTIVE PHOTOPRODUCTION IN DIS}
\author{I. Royzen }
\begin{document}
\maketitle
\centerline{\it {Lebedev Physical Institute, 53 Leninski Prospect,
117333 Moscow, Russia}}
\addtolength{\baselineskip}{8pt}
\begin{abstract}
A simple relativistic model is suggested that elucidates
qualitatively the quark-antiquark distribution within virtual photon.
The diffractive hadroproduction in DIS initiated by highly virtual photon
$\gamma ^{*}(Q^2)$ is discussed in more detail. The main result is that the
contribution of
large transverse scale $q\bar q$-fluctuations of the photon is just sufficient
to produce the cross section of its hadron-like strong interaction
(in particular, its diffraction into hadrons) of the same $Q^2$-dependence as
the total cross section of $\gamma ^{*}p$-interaction. It is why observed in
DIS fraction of photon diffractive hadroproduction in \enspace
$\gamma^{*}p$\enspace interaction is quite large and does not vary with $Q^2$.
\end{abstract}
\vspace{7mm}
 
The unambiguous experimental evidence has been obtained ~\cite{aa} in DIS
that the cross section of photon diffraction into hadrons in the
process \enspace $\gamma^{*}(Q^2)+p\rightarrow hadrons$ \enspace is unexpectedly
large even at
$Q^2\gg 1\mbox{ $GeV$}^2$, and that its fraction in the total cross section of
\enspace $\gamma ^{*}p$\enspace interaction is almost independent of $Q^2$.
The data obtained at HERA, using the H1 detector,
as well as their theoretical interpretation in terms of the proton diffractive
structure function were summarized in ref. ~\cite{pn}.
Meanwhile, being of essentially nonperturbative origination, these results
unavoidably imply the significant role which play fluctuations of highly
virtual space-like photon into hadronic states of sufficiently large transverse
size. That is why the analysis in terms of the photon structure function seems
to be more adequate. The account of rare fluctuations of such a type was shown
would like to present a very simplified two-particle model which demonstrates
that the mean transverse size of highly virtual space-like photon treated as a
function of two variables, $Q^2$ and $x$ ($x$ being the familiar Feynman
variable - energy fraction carried by (anti)quark), is really expected to peak
sharply at sufficiently small \enspace$x\sim Q^{-2}$\enspace, having {\em
{irrespectively of $Q^2$}} its maximum value about $1\enspace GeV^{-1}$. No
doubt, these rare large scale parton configurations of the photon
\enspace$\gamma^{*}(Q^2)$\enspace should produce the diffraction pattern
similar to that observed in the scattering of the real photon (vector mesons).
Since the width of the peak is shown to be proportional to $Q^{-2}$, one can
easily understand that their contribution to the cross section
of\enspace$\gamma^{*}p$ interaction exhibits nearly the same
\enspace$Q^2$-dependence as the bulk
of short-range ("point-like")\enspace$\gamma^{*}p$ collisions. Being the
striking
qualitative effect, this feature of virtual photon interaction is expected
to manifest itself also within the frameworks of more realistic approaches.
 
In the present communication, the $q\bar q$-distribution within space-like
photon is linked
to the quark structure of time-like hadronic states in the relevant channel.
No unreliable perturbation theory calculations are invoked, however the
nonperturbatively motivated treatment is paid by implication of some specific
model assumptions.
 
The space-like virtual photon state is assumed to be related to
the time-like $q\bar q$-states in the conventional quantum
mechanical manner:
the contribution of a given state is proportional to the (lightcone) time
$\Delta{t}$ which
this state "spends", having the energy of the virtual photon $\gamma^{*}(Q^2)$
(or vice versa), in the accordance with the energy-time uncertainty relation.
The states to be allowed for incorporate
the low laying resonances ($\rho _0,\omega ,etc.$) as well as the continuum
background which contributes predominantly to the parton distribution at
large $Q^2$. It is easy to estimate that for the state with the eigenmass
$M \gg 1\,GeV$ and momentum $\vec P$\enspace($|\vec P|/M\enspace\equiv P_z/M
= \gamma \to \infty$) \begin{eqnarray}\label{1} \Delta{t}\enspace
\sim\enspace \frac{1}{\sqrt{{\vec P}^2 + M^2}-\sqrt{{\vec P}^2 - Q^2}}\enspace
\simeq\enspace \frac{2\vert{\vec P}\vert}{M^2 + Q^2} \end{eqnarray} Hence, if
a certain wave function can be attributed to space-like $q\bar
q$-fluctuation of four momentum squared equal to $-Q^2$, then the coefficients
in its decomposition into the series (integral) over
the eigenfunctions of time-like states are reasonably expected to be
proportional to $(M^2 + Q^2)^{-1}$, and therefore, the parton distribution in
the time-like state of mass $M$ contributes to that in the above space-like
fluctuation with the weight proportional to $(M^2 + Q^2)^{-2}$.
 
As a very simple realization of this guess,
one can consider the two-particle model of spinless quark and antiquark of mass
$m_q$, interacting via the potential $U$ of the simplest form: $U =
0$\enspace within the relativistically contracted (along the axis $z$)
ellipsoid of transverse radius $R$ about some few $GeV^{-1}$ and $U = \infty$
outside it.\footnote{The many particle (evolution) aspect of the problem is
briefly discussed below.} Since the highly excited states,
\enspace$4m_q^2 \ll M^2\sim Q^2$ \enspace are of primary significance at large
$Q^2$, one can easily anticipate that what follows should be almost independent
of the precise shape of potential within this ellipsoid, provided that the
effectively non-transparent potential wall near its surface exists: being far
above, these states do not "feel" the peculiarities of quark interaction therein.
Of course, being associated with
peculiarities of the potential $U$, quantization of the mass $M$ is
irrelevant too.
 
The corresponding relativistic Shr\H odinger equation
\begin{eqnarray}\label{2} \lbrace\enspace \sqrt{{\vec p}^2_1 +m^2_q} +
\sqrt{{\vec p}^2_2 + m^2_q} + U - \gamma M \enspace\rbrace\enspace{\psi(\vec
p_1,\vec p_2)}\enspace=\enspace 0 \end{eqnarray} should be supplemented by the
condition \begin{eqnarray}\label{3} \vec p_1 + \vec p_2 = \vec P \end{eqnarray}
The solution (more precisely, its absolute value) of Eq.(2) is
obviously peaked near its classical limit:
$$\sqrt{{\vec p}^2_1 +m^2_q} +\sqrt{{\vec p}^2_2 + m^2_q} -
\gamma M = 0 $$ For the highly excited states, $M \gg 2m_q$, which are just
relevant, the quasiclassical approximation should be valid, i.e., the width
of quantum distribution of particle momenta around this strict
constraint is relatively small \footnote{The exactly solvable example
in support of the following estimate.}
:$${\left\vert\frac{\sqrt{{\vec p}^2_1 +m^2_q} +\sqrt{{\vec p}^2_2
+ m^2_q}}{\gamma M} - 1\right\vert} \simeq \frac{\pi}{RM} \ll 1$$ That is why in
the context of the following discussion, the precise solution of Eq.(2) can be
reasonably replaced by \begin{eqnarray}\label{4} \psi(\vec p_1,\vec p_2) \sim
\delta (\sqrt{{\vec p}^2_1 + m^2_q} + \sqrt{{\vec p}^2_2 + m^2_q} - \gamma M)
\end{eqnarray}
 
At ${\vec P}^2 \gg M^2$ this solution describes the eigenstates of the
angular momentum $\mathaccent94 J$ (in particular, $J = 1$) as well (for the
moment, the normalization is put aside) because in the above limit the
approximate relations\enspace $|\vec p_1|\simeq p_{1z}$,\enspace $|\vec
p_2|\simeq p_{2z}$ and\enspace $J\simeq J_z$ are maintained ~\cite{dd}, and
therefore, operator $\mathaccent94 J$ is essentially commutative with what is
enclosed in the curly brackets of Eq.(2).  Nevertheless, the selection of
states with $\mathaccent94 J = 1$ crucially affects the subject: since the
radial excitations are in life only, the states of different masses $M$
contribute to the integral over $M$ with the same (independent of $M$) weight.
Just because of this restriction, no factor of the type \,
$\sum_{J=0}^{M/M_0}(2J+1)$ \,appears in the integrand in the middle of Eq (6).
 
Making use the well known relativistic kinematics, one can replace the
variables in Eq.(4) by the more convenient ones: Feynman variable
$x \simeq p_{1z}/(p_{1z} + p_{2z}) \equiv p_{1z}/|\vec P|$ and, so-called,
transverse mass \,$m_T = \sqrt{p^2_T + m^2_q}$\,
where $p_T$ is the absolute value of quark transverse momentum. Being expressed
in terms of new variables, Eq.(4) reads\\ (as $\gamma \to \infty$)
\begin{eqnarray}\label{5} \psi \enspace\sim \enspace\delta
(\frac{m_T}{\sqrt{x(1 - x)}} - M) \end{eqnarray} Thus, for the quark
distribution $dW$ in a virtual state with $Q^2\gg 1\enspace GeV^2$ one gets $$
\int\frac{dW}{d^3{\vec p}_1d^3{\vec p}_2}\delta^{3}(\vec p_1 + \vec p_2 - \vec
P) d^3(\vec p_1 +\vec p_2)\enspace\sim\enspace$$ \begin{eqnarray}\label{4}
{\left\vert\int\limits_{2m_q}^{\infty}\frac{\delta(m_T/\sqrt{x(1 -
x)} - M)dM}{M^2 + Q^2}\right\vert}^2 \enspace=\enspace\frac{1} {[Q^2 + m^2_T/x(1
- x)]^2}\end{eqnarray}   The standard relativistic kinematics gives (as $\gamma
\to \infty)$ $$ d(p_{1z}-p_{2z})\enspace\simeq
\enspace\frac{\gamma m_Tdx}{2x^{3/2}(1-x)^{3/2}}$$ and, finally, the
normalized (and integrated over the azimuthal angle) quark distribution is
obtained in the form
:  \begin{eqnarray}\label{7} \frac{dW}{dm_Tdx}\enspace
\simeq\enspace {\frac {4Q}{\pi}} \frac{m^2_T} {x^{3/2}(1 - x)^{3/2} \lbrack Q^2
+ m_T^2/x(1 - x)\rbrack^2} \end{eqnarray} This distribution contains much more
information on the object, than the familiar inclusive structure function
itself (the latter one is obtained by integration of Eq.(7) over $m_T$ which
results obviously in the loss of some important details: one can easily check
that it is almost perfectly plateau-like except of very narrow intervals ($\sim
m_q^2/Q^2$) near $x = 0$ and $x = 1$ where it tends to zero).  The most
essential feature of this distribution shows up, if one estimates the mean
transverse size of the virtual $q\bar q$-fluctuation as the function of
variables $x$ and $Q^2$ \begin{eqnarray}\label{8}<r(x,Q^2)>
\enspace\simeq\enspace <\frac{1}{m_T}>\enspace\simeq  \int
\limits_{m_q}^{\infty} {\frac{dm_T}{m_T}
\frac{dW}{dm_Tdx}}\enspace\simeq\enspace \frac{1}{\pi
m_q}\enspace\frac {2y}{1 + y^2}\end{eqnarray} where
$$y = \frac{Q\sqrt{x(1 - x)}}{m_q}$$
Thus, the mean transverse size of $q\bar q$-state \enspace
$<r(x,Q^2)>$\enspace is peaked sharply at
$$x(1 - x) = \frac{m^2_q}{Q^2}$$
i.e., very close to $x = 0$, see Fig.1 (in the framework of the oversimplified
two-particle model under consideration the peak near
$x = 1$ is nothing else, than again the above one only associated with
the second parton).
The main point is that its maximum value is equal to
\begin{eqnarray}\label{9} r_{max}\enspace \simeq\enspace (\pi m_q)^{-1}\enspace
\geq\enspace 1 \mbox{ $GeV$}^{-1},\end{eqnarray}
{\em irrespectively of \enspace$Q^2$},\enspace being about
\enspace$Q/4m_q$\enspace times larger, than its typical
"point-like" expectation (i.e., than its value at $x \simeq 0.5$). One can
estimate the value of $r_{max}$ under the well known assumption that the light
quark masses are proportional to the distance between them, $m_q \simeq Kr$,
where $K \simeq \frac{1}{2\pi}\mbox{ $GeV$}^{-2}$. Then, the typical scale of
large-size $q\bar q$-fluctuations is $r_{max} \simeq \sqrt{2}\mbox{
$GeV$}^{-1}$ (it is worthy to note that this scale is just about correlation
radius for perturbative gluons motivated by the lattice studies ~\cite{sh}),
and corresponding quark mass $m_q \simeq Kr_{max} \simeq 220 \mbox{ MeV}$.
Since it is only slightly less, than the mass of constituent quark (valon), the
effective radius of quarks in large-size $q\bar q$-fluctuations is just of the
order of (or slightly less, than) that of valons which was estimated ~\cite{an}
to be about (1.0 - 1.8)\enspace $GeV^{-1}$. Thus, the large size $q\bar
q$-fluctuations just make quark and antiquark spatially resolvable from each
other, both color and electric dipoles becoming recognized and, hence, strong
and electromagnetic interactions between $q$ and $\bar q$ coming in life.
 
The considered $q\bar q$-fluctuations can be reasonably related to the virtual
photon $\gamma ^{*}(Q^2)$ and its interaction with photon as it is shown in
Fig. 2. The account of small size fluctuations ($q\bar q,\enspace e^{+}e^{-},
\enspace etc.)$, $r\sim Q^{-1}$, is nothing else, than the dual ("microscopic")
way of description of the "ordinary point-like" photon (in a sense, one can
even say that
they are not fluctuations at all), and perturbative treatment of these
fluctuations is, at least, doubtful (the equivalence of both representations is
emphasizes in Fig.  2(a) by matching them with an effective coupling constant
about 1), while the large size (true) fluctuations, $r\sim r_{max}$, are
coupled to the photon by the electromagnetic constant $\alpha$ but show up for
it the hadron-like strong interaction with proton (an effective coupling
constant about 1 in Fig.  2(b)).  That is why {\em in the process $\gamma ^{*}p
\to hadrons$} the virtual photon can be essentially thought of as an object
described by the above $q\bar q$-distribution and coupled to the
proton by the constant $\alpha$. 
\footnote{This distribution resembles, to some extent,
perturbatively motivated distributions ~{\cite{nn},\cite{hd}} although it is
far from being identical with them.}
 
The large size $q\bar
q$-fluctuations should show up {\em{the hadron-like strong
interaction}} with the cross section which is caused by the degree of color
descreening, i.e., is of the order of \enspace$\pi B(r)<r^{2}(x,Q^2)>
\enspace\simeq\enspace 0.6B(r)/m_q^2$ where
$B(r)$ is some unknown phenomenological non-transparency (blackness) factor,
$0 = B(0) < B(r_{max}) \simeq 1$. Alongside with this, the probability of such
fluctuations decreases from 1 to $\alpha$ due to the electrical charge
descreening (see above).
To get an estimate one can take $B(r) \equiv
0$ and the fluctuation probability equal to 1 at $r < r(0)$ (the
electromagnetic "point-like" $\gamma^{*} p$ interaction shows up only) and
$B(r) \simeq 1$ within the range $r_0\leq r\leq r_{max}$, the $q\bar
q$-fluctuation probability being proportional to $2\pi\alpha$ (strong $\gamma^{*}
p$ interaction is dominated).  Then, being integrated over $x$, the above cross
section of hadron-like $\gamma ^{*}p$ interaction is obviously proportional to
$Q^{-2}$ (the width of the dashed domain in Fig.1), and therefore, its ratio to
the cross section of the "point-like" \enspace$\gamma^{*}p$\enspace interaction
should be nearly independent of $Q^2$. It is easy to estimate
that these two contributions \footnote{The interference between corresponding
amplitudes is undoubtedly unessential because of qualitatively different
mechanism of the final state generation.} are nearly equal at quite reasonable
value of $r_0$, $r_0 \simeq 0.9\enspace r_{max} \simeq 1.25\enspace GeV^{-1}$.
At the same time, the part of the virtual photon interaction associated with
its hadron-like fluctuations should obviously show up the features of the real
photon one.  In this case the fraction of virtual photon diffractive
hadroproduction observed in DIS could be easily understood, if one adopts that
it is about 20 \% of the total cross section caused by its hadron-like
interaction (just like for the real photon).
 
Of course, for being more realistic, the considered model needs some essential
improvements.  Nevertheless, its striking qualitative
features can, most probably, stand against the necessary modifications which
are of two kinds.  First, even in the framework of the two-particle
approximation one has to allow for the direct quark-antiquark interaction via
gluonic exchange that should result in formation of an effective (linear in
the "distance"  $\sqrt{4R^2 + \gamma ^2 (z_1 - z_2)^2}$\enspace) confining
potential. As it was already mentioned, the
relevant corrections do not affect noticeably the above results.  Second, the
many particle aspect of the problem is to be taken into account.  In the spirit
of the suggested approach it should be equivalent to the replacement of the
above artificial non-transparent potential wall and corresponding infinite set
of excited two-quark (time-like) states by the real many particle (quark and
gluon) state continuum. One can reasonably anticipate that energy dissipation
over the many degrees of freedom (at the same $M$) is only favorable for
appearing the low $x$ partons and, consequently, for realization of the large
size states. In terms of perturbation theory one can say that evolution and
gluon bremsstrahlung (especially important is the latter one produced by the
low $x$ quark or antiquark) enlarge the population of low energy partons and
could lead, most probably, to an enhancement of the large scale effects.
Anyway, the distributions (7) and (8) can be used as quite reasonable starting
points for the calculation of more realistic parton distribution within highly
virtual space-like photon. In particular, they can be chosen as the initial
conditions which are well known to influence significantly the result of
solution of the evolution equation.  At the same time, the above discussions
provides some insight to the essential complications that are to be faced in
the program of usage of DIS as an effective tool for investigation of the
short-range structure of hadrons.
 
The work is supported by Russian Foundation for Basic Researches, Grants
No. 96-02-16347 and No. 96-02-19572.\\

\end{document}